\newif\iflayout
\layouttrue

\iflayout
   \documentclass[twocolumn,prb]{revtex4}
\else
   \documentclass[preprint,prb]{revtex4}
\fi

\usepackage{epsfig,amssymb}
\usepackage{bm}

\begin{document}

\title{First-principles calculations of spin spirals in Ni$_2$MnGa and Ni$_2$MnAl}
\author{J. Enkovaara}
\email{jen@fyslab.hut.fi}
\author{A. Ayuela}
\author{J. Jalkanen}
\affiliation{Laboratory of Physics, Helsinki University of Technology}
\author{L. Nordstr\"om}
\affiliation{Condensed Matter Theory, Uppsala University}
\author{R.M. Nieminen}
\affiliation{Laboratory of Physics, Helsinki University of Technology}

\begin{abstract}
We report here non-collinear magnetic configurations in the Heusler alloys
Ni$_2$MnGa and 
Ni$_2$MnAl which are interesting in the context of the magnetic shape memory
effect. The total energies for different spin spirals are calculated and the 
ground state magnetic structures 
are identified.
The calculated dispersion curves are used to estimate the
Curie temperature which is found to be in good agreement with experiments.
In addition, the variation of the magnetic
moment as a function of the spiral structure is studied. Most of the
variation is associated with Ni, and symmetry constraints relevant for the
magnetization  
are identified. Based on the calculated results, the effect of the constituent
atoms in determining the Curie temperature  
is discussed.

\end{abstract}


\maketitle

\unboldmath

\section{Introduction}

Materials showing strong coupling between the magnetic and structural
properties are interesting from the technological point of view. Tb-Dy-Fe
alloys 
(Terfenol-D, already in commercial use ) exhibit magnetic field induced
strains of $\sim$~0.1~\% based  
on the magnetostriction phenomenon \cite{Clark80}. On the other hand, Ni-Mn-Ga
alloys  
close to the Ni$_2$MnGa stoichiometry show strains up to 10~\% with moderate
magnetic fields \cite{Murray00, Heczko00, Sozinov02}. The mechanism of this
phenomenon, the magnetic shape memory (MSM) effect, is based on the
magnetic-field-induced movement of structural domains (twin variants)  and is 
different from ordinary magnetostriction \cite{RJames00}. 
The basic magnetic
properties related to the MSM effect include the saturation magnetic moment
and the 
magnetic anisotropy which have been studied earlier for Ni$_2$MnGa
\cite{Ayuela02,Enkovaara02}. Here, we probe deeper into
the magnetic properties of Ni$_2$MnGa and an other MSM candidate, Ni$_2$MnAl,
by 
studying 
non-collinear magnetic configurations which also enables one to  consider 
finite 
temperature effects in a natural way.  

\begin{figure}[h]
\centering \epsfig{file=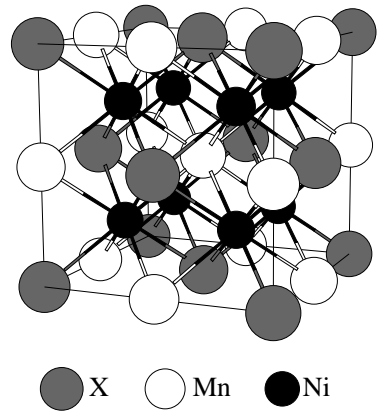,width=.6\columnwidth,keepaspectratio=true}
\caption{Cubic cell of the L2$_1$ structure, where X is Al or Ga. The cubic
cell contains four primitive cells.}
\label{fig:l21}
\end{figure}

Although one ingredient in the MSM effect is a structural transformation
(martensitic 
transformation) from a cubic structure to a lower-symmetry
structure upon cooling, we concentrate here only in the high temperature
phase.
In this phase Ni$_2$MnGa has the cubic L2$_1$
structure (see Fig.~\ref{fig:l21}) as shown by x-ray and neutron diffraction
measurements \cite{Webster69,WebsterEtal84}. The magnetic order is
ferromagnetic and most of the magnetic moment originates from Mn
\cite{WebsterEtal84,Ayuela02}. In the stoichiometric compound the Curie
temperature is about 370 K\cite{WebsterEtal84} and decreases when
increasing the Ni content \cite{Vasilev99}. On the other hand, Ni$_2$MnAl
is less studied and its 
structure and magnetic configuration do not seem to be perfectly
understood. On the structural side, both 
L2$_1$ and disordered B2 structures are reported
\cite{ZiebeckWebster75,Morito98,Fujita00,acet02,manosa02} depending on the
thermal 
treatment. The magnetic configuration is found to be ferromagnetic with Curie
temperatures between 300 K and 400 K in Ref. \onlinecite{Fujita00} and
antiferromagnetic or spiral in Refs. \onlinecite{ZiebeckWebster75,Morito98}. The
magnetic moment comes mainly from Mn atoms also in this compound 
\cite{ZiebeckWebster75,Ayuela99}.
It seems that the ground state magnetic configuration depends on
the underlying crystal structure. Here we address the possibility of
non-collinear magnetic configurations in the L2$_1$ structure.

Although the original formulation of the local-spin-density approximation
\cite{vonBarthHedin72} of density-functional theory allowed non-collinear
magnetic 
order, first-principles calculations for this aspect have started only recently
(for a review see Ref. \onlinecite{Sandratskii98}). One application has been the
study of non-collinear ground states for example in $\gamma$-Fe
\cite{Sandratskii92,Bylander99,Knopfle00} or in frustrated antiferromagnets
\cite{Kurtz01,Hobbs00}.  
In addition, the non-collinear formulation
enables studies of finite temperature properties of magnetic materials.
As the dominant magnetic excitations at low temperatures are spin waves which
are non-collinear by nature, it is possible to determine the 
magnon spectra and ultimately the Curie temperature from first principles
\cite{Rosengaard97,Uhl96,Halilov98,Pajda01}.
Most of the previous work has
been done for elements or compounds with only one magnetic constituent. We
study here systems with several magnetic atoms and show how the interaction
between different magnetic sublattices can give rise to interesting effects.

The paper is organized as follows.
Some general properties of spin spirals are discussed in Sec.~\ref{sec:spiral}
followed by the description of the computational scheme in
Sec.~\ref{sec:comp}. We study the total energy and magnetization with spiral
magnetic orderings and estimate the Curie temperature in
Sec.~\ref{sec:results}
and finally we conclude in Sec.~\ref{sec:conclusions}.

\section{General properties of spin spirals}
\label{sec:spiral}

The magnetic configuration of an incommensurate spin spiral shows the 
magnetic moments of certain atomic planes varying in direction. The variation
has a 
well-defined period 
determined by a wave vector $\bm{q}$. When the magnetic moment is
confined to the lattice sites the magnetization $\bm{M}$ varies as
\begin{equation}
\bm{M}(\bm{r}_n)=m_n \left( \begin{array}{c}
\cos(\bm{q} \cdot \bm{r}_n + \phi_n) \sin(\theta_n) \\
\sin(\bm{q} \cdot \bm{r}_n + \phi_n) \sin(\theta_n) \\
\cos(\theta_n) \end{array} \right),
\label{eq:spir1}
\end{equation}
where polar coordinates are used and $m_n$ is the magnetic moment of atom $n$
with a phase $\phi_n$ at the position $\bm{r}_n$. Here, we consider only
planar spirals, that is 
$\theta_n=\pi/2$ which also gives the minimum of the total energy. The
magnetization of Eq.~(\ref{eq:spir1}) is not 
translationally invariant but transforms as
\begin{equation}
\bm{M}(\bm{r}+\bm{R})= D(\bm{q}\cdot \bm{R}) \bm{M}(\bm{r}),
\label{eq:spir2}
\end{equation}
where $R$ is a lattice translation and $D$ is
a rotation 
around the $z$-axis. A spin spiral with a magnetization in a general point
$\bm{r}$ in space  can be defined as a magnetic
configuration which transforms according to Eq. (\ref{eq:spir2}). As the spin
spiral describes a spatially 
rotating magnetization, it can be correlated with a frozen
magnon. 

Because the spin spiral breaks the translational symmetry, the Bloch
theorem is no more valid. Computationally, one should use large supercells to
obtain total energies spin spirals. 
However, one can define generalized
translations which contain translations in real space and rotations in spin
space \cite{Herring66,Sandratskii91}. These generalized
translations leave the magnetic structure invariant and lead to
a generalized Bloch theorem. Therefore the Bloch spinors can 
still be characterized by a $\bm{k}$-vector in the Brillouin zone, and 
can be written as
\begin{equation}
\psi_k(\bm{r})= e^{i \bm{k} \cdot \ \bm{r}} \left( \begin{array}{l}
e^{-i \bm{q}  \cdot \bm{r}/2} u_k(\bm{r}) \\
e^{+i \bm{q}  \cdot \bm{r}/2} d_k(\bm{r}) \end{array} \right).
\end{equation}
The functions $u_k(\bm{r})$ and $d_k(\bm{r})$ are invariant with respect
to lattice translations having the same role as for normal Bloch
functions. Due to this  
generalized Bloch theorem the spin spirals can be 
studied within the chemical unit cell and no large supercells are needed.

Although the chemical unit cell can be used, the
presence of the spin spiral lowers the symmetry of the system. Only the
space group operations that leave invariant the wave vector of the spiral
remain. When considering the general spin space groups, i.e. taking the spin rotations
into account, the space group
operations which reverse the spiral vector together with a spin rotation of $\pi$
around the x-axis are symmetry operations \cite{Sandratskii91}.

Basically, the spin spiral relates only the magnetizations in the different
primitive cells. However, the symmetry properties constrain
the magnetization which we discuss here in the context of the L2$_1$
structure. The primitive cell of the L2$_1$ structure (one fourth of the cubic
cell shown in Fig.~\ref{fig:l21}) contains four atoms, two
Ni, one Mn and one Ga or Al atom.  In the full cubic symmetry the two Ni
atoms are equivalent but this equivalence can be broken when the spin spiral
lowers the symmetry of the system. If the 
spiral wave vector is in the [111] direction the two Ni atoms are
no longer equivalent under space group operations. Considering also the spin
rotations,  the phases $\phi_n$ of
the two Ni magnetizations are opposite as the atoms are related by space
inversion. 
If the two Ni atoms are treated equivalent (when allowed by the spiral 
symmetry) are constrains for the phases of Ni moments even stronger.
If the magnetic moments of Ni within the
primitive cell are $\bm{M}(\bm{r}_1)=m_1 cos(\phi_1)=\bm{M}(\bm{r}_2)$, the
magnetic moment in the 
neighbouring cell at $-\bm{r}_1$ is $\bm{M}(-\bm{r}_1)=m_1 cos(-\phi_1)$. On
the other hand, 
the Ni atoms at $-\bm{r}_1$ and at $\bm{r}_2$ are connected by a lattice
translation, so 
that according to Eq.~(\ref{eq:spir2}) $\bm{M}(\bm{r}_2)=m_1 cos(-\phi_1 + \bm{q} \cdot \bm{R})$
and one has the relation $\phi_1=-\phi_1 + \bm{q} \cdot \bm{R}$ for the
phase. In order to obtain the true minimum energy configuration it may be
necessary to treat the Ni atoms as inequivalent (i.e. lower the symmetry of
the system) so that the above relation for the phase do not have to hold.

\section{Computational method}
\label{sec:comp}

The spin spirals discussed in section~\ref{sec:spiral} are studied within
the density-functional 
theory. We use the full-potential linearized augmented-plane-wave method
(FLAPW) method \cite{Wimmer81} in an implementation which allows non-collinear
magnetism including spin spirals \cite{Nordstrom96,Nordstrom00}. In addition
to the full 
charge density and to the full potential, the full magnetization density is
used. The magnetic moment is allowed to vary both in magnitude and in
direction inside the atomic spheres as well as in the interstitial regions.
The plane wave cut-off for the 
basis functions is $RK_{max}=9$, leading to $\sim$ 350 plane waves with the
muffin-tin radii 2.25 
a.u. Brillouin zone integrations are 
carried out with the special point method using 800 $k$-points in the full
Brillouin zone and a Fermi broadening of 0.005 Ry. For the
exchange-correlation potential we use both the local-spin-density  
approximation (LSDA) \cite{vonBarthHedin72} and the generalized
gradient approximation (GGA) \cite{PerdewWang92} which we discuss next in more
detail.

\subsection{LSDA vs GGA}

It has been pointed out that the use of GGA is beneficial in the context of
Ni$_2$MnGa \cite{Ayuela99,Ayuela02}. Because there has been some
discussion about the different exchange-correlation 
potentials in the context of non-collinear magnetism, we present some
comparison also here. 

Although there is no global spin quantization axis, one can consider at
every point of space a local coordinate system such that the magnetization at
that point is in the z-direction. As the LSDA depends only
on the magnitude of the magnetization, the exchange-correlation
potential can be 
calculated at every point in the local coordinate system as in the usual
collinear case. The non-collinear potential is obtained by rotating back
to the global frame of reference. On the other hand, the 
GGA depends also on the gradients of magnetization. Because the magnetization
direction may vary, only projections of the magnetization on the local
quantization axis are used in the standard GGA when evaluating the
gradients. If the magnetization direction varies slowly this should not bring
any problems. Some previous work has suggested  that the
disagreements between theory and 
experiment are due to projection errors in some cases
\cite{Bylander99}. However, later work has 
corroborated that the main issue is not the exchange-correlation functional but
the actual computational method, pointing to the importance of all-electron
and full-magnetization treatment \cite{Knopfle00,Sjostedt02,Hafner02}.

\begin{figure}[h]
\centering \epsfig{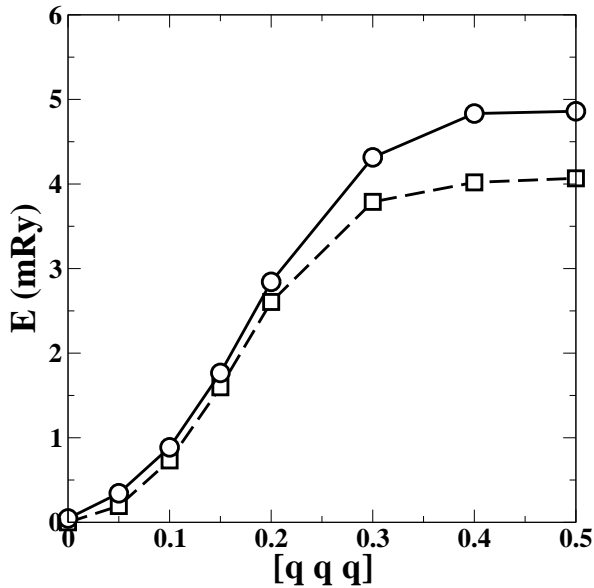}
\caption{Total energy as a function of the spiral vector
 $\bm{q}$ in units of $2\pi/a$. $\bigcirc$
LSDA, $\square$ GGA.} 
\label{fig:lsda}
\end{figure}

We have done all the calculations in this work
 both with LSDA and with
GGA. The
total energy as a function of the spiral wave vector length  in Ni$_2$MnGa is
 shown in Fig. \ref{fig:lsda} for a single direction. 
One can see that for small $q$ both approximations give 
similar 
results. With larger $q$ the results differ slightly but the same
qualitative 
behaviour is seen. 
For the other results presented in the following sections the qualitative
behaviour is also the 
same for LSDA and GGA, and the quantitative differences between the two
approximations 
are even smaller. Therefore, only the  GGA results are discussed in the
following.

\section{Results and discussion}
\label{sec:results}

\subsection{Total energies}

\begin{figure}[h]
\centering \epsfig{file=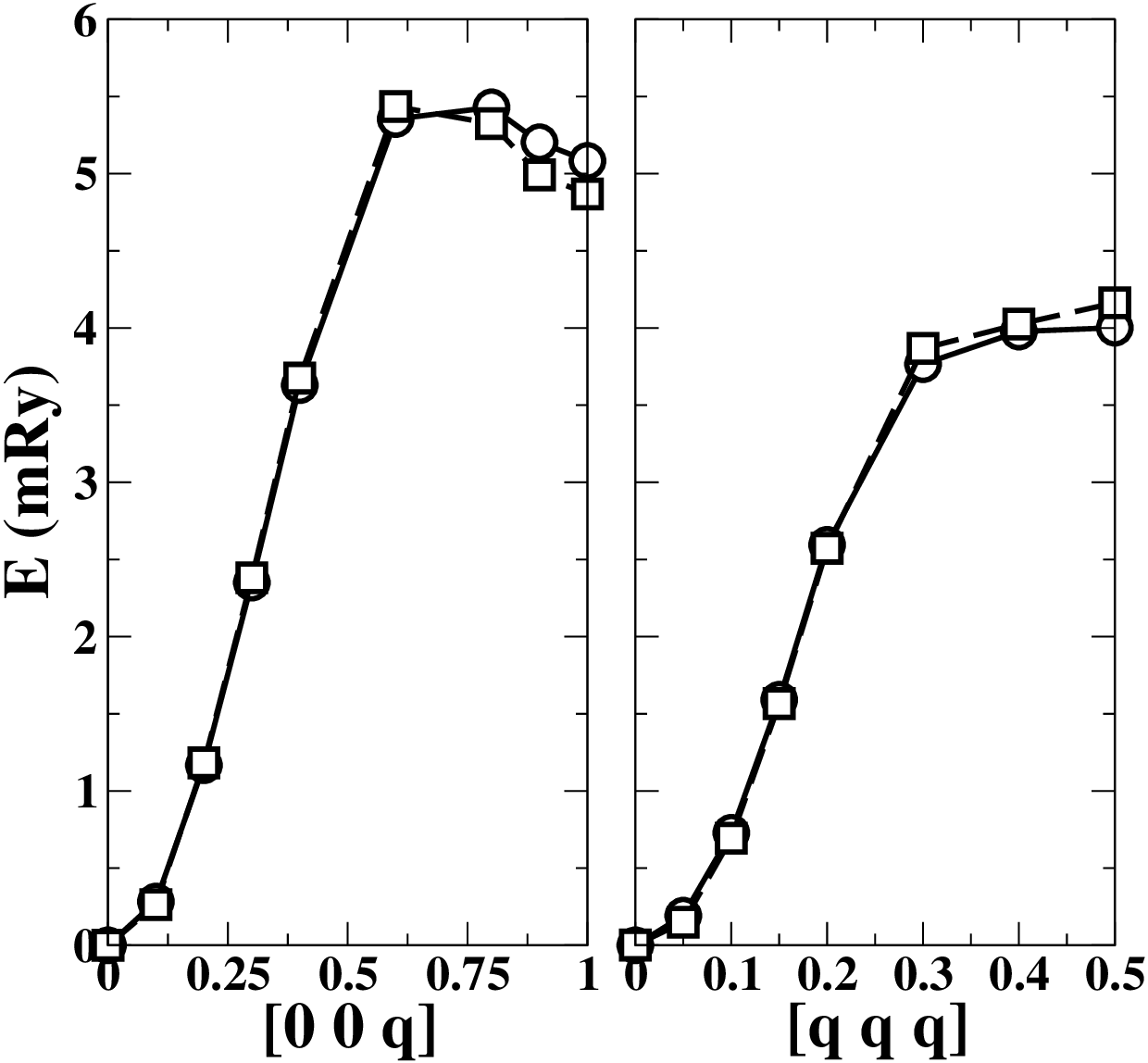,width=.9\columnwidth,keepaspectratio=true}
\caption{Total energy as a function of the spiral vector $\bm{q}$. $\bigcirc$ Ni$_2$MnAl, $\square$ Ni$_2$MnGa.}
\label{fig:ene1}
\end{figure}

\begin{figure}[h]
\centering \epsfig{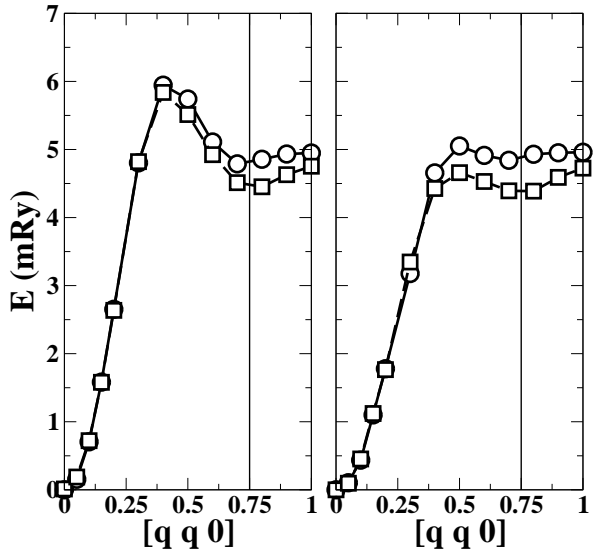}
\caption{Total energy as a function of the spiral vector $\bm{q}$. $\bigcirc$
Ni$_2$MnAl, $\square$ Ni$_2$MnGa. a) Ni atoms are equivalent, b) Ni atoms
are inequivalent. Vertical line denotes the Brillouin zone boundary.}  
\label{fig:ene2}
\end{figure}

First, we have studied the possibility of non-collinear ordering by studying
the energetics of spiral configurations. This study
also provides information about finite-temperature properties.
The total energy is calculated as a function of the spiral wave
vector $\bm{q}$, and the wave vector is varied along the high symmetry
directions 
$[001]$, $[110]$ and $[111]$. $\bm{q}$ is given in units
of $2 \pi /a$ where $a$ is the theoretical lattice constant of the 
L2$_1$ structure \cite{Ayuela99}. The corresponding total energies are shown
in Figs.~\ref{fig:ene1}~and~\ref{fig:ene2}. 

Fig.~\ref{fig:ene1} shows that the variation of total
energy in $[001]$ and $[111]$ directions is similar in Ni$_2$MnGa and
Ni$_2$MnAl for all values of  $\bm{q}$. The lowest energy in all cases is at
$q=0$ which is the normal 
collinear ferromagnetic configuration. 
Both materials have  small minimum at the antiferromagnetic configuration at
$\bm{q}=(0\ 0\ 1)$, but at other antiferromagnetic configuration at
$\bm{q}=(0.5\ 0.5\ 0.5)$ there are no minima.

The energy in the [110] direction is also similar for both materials as seen in
Fig.~\ref{fig:ene2}. Here, the effect of symmetry constraints can be seen
clearly. If the two Ni atoms are equivalent the energy is higher especially
around $\bm{q}=(0.5\ 0.5\ 0)$. When the magnetic moments of the
two Ni are allowed to relax 
independently the energy lowers and the dispersion becomes flat after
$\bm{q}=(0.5\ 0.5\ 0)$. Near the Brillouin zone boundary at
$\bm{q}=(0.75\ 0.75\ 0)$ both materials show small energy minima corresponding to
incommensurate spiral order. At the antiferromagnetic configurations at
$\bm{q}=(1\ 1\ 0)$ there are no clear energy minima even though in the case of
Ni$_2$MnAl the dispersion is very flat.

Generally, the spin spirals are related to magnons which allows 
the estimation of magnon-related properties, such as spin stiffness and
Curie 
temperature, from the total energies
calculated above. 
The total energy of the planar spin spiral is related to the magnon energy
$\omega_q$ 
as \cite{Rosengaard97,Halilov98}
\begin{equation}
\omega_q=\frac{4 \mu_B}{M}E(\bm{q}),
\end{equation}
where $M$ is the magnetic moment per unit cell.
In the low $q$ limit the magnon dispersion is quadratic, and one defines the
spin 
stiffness constant $D$ as
\begin{equation}
\omega_q=D q^2. \label{eq:stiff}
\end{equation}
From the calculated total energies in
Figs.~\ref{fig:ene1}~and~\ref{fig:ene2} we can estimate the same spin
stiffness for both materials 
which is $D=77$ mRy a.u.$^2$. 
This in good agreement with the experimental value 79~mRy a.u.$^2$ measured in 
Ni-Mn-Ga films \cite{Patil02}.

The Curie temperature can be estimated on the basis of the Heisenberg model.
By mapping the first-principles results to the Heisenberg model, the Curie
temperature $T_C$ in the random phase approximation is given by
\cite{Pajda01,Wang82} 
\begin{equation}
\frac{1}{k_B T_C} = \frac{6 \mu_B}{M}\frac{V}{(2\pi)^3}\int d^3q \frac{1}{\omega_q},
\end{equation}
where $V$ is the unit cell volume,
and the integration is over the Brillouin zone. An estimation can be obtained
using the quadratic dispersion, Eq.~(\ref{eq:stiff}) and carrying out the
integration 
over a sphere having the same volume as the Brillouin zone. This results in
\begin{equation}
\frac{1}{k_B T_C}=\frac{3 V q_d}{M \pi^2 D},
\end{equation}
where $q_d=(6 \pi^2 /V)^{1/3}$. By using the calculated spin stiffness
constant we obtain $T_C$=830 K which is clearly an overestimate. As seen in
Figs.~\ref{fig:ene1}~and~\ref{fig:ene2} the dispersion curve $E(\bm{q})$ deviates strongly from
the 
quadratic behaviour with larger $q$. A better estimate can be obtained
by considering the dispersion quadratic up to some radius and constant
thereafter. Based on the calculated energies in Figs.~\ref{fig:ene1}~and~~\ref{fig:ene2} the
constant is chosen to be 5 mRy 
when $q>0.7 q_d$. The Curie temperature obtained
in this way is $T_C$=485~K which compares well with the experimental one
380~K.  

\subsection{Magnetic moments}

\begin{figure}[h]
\centering \epsfig{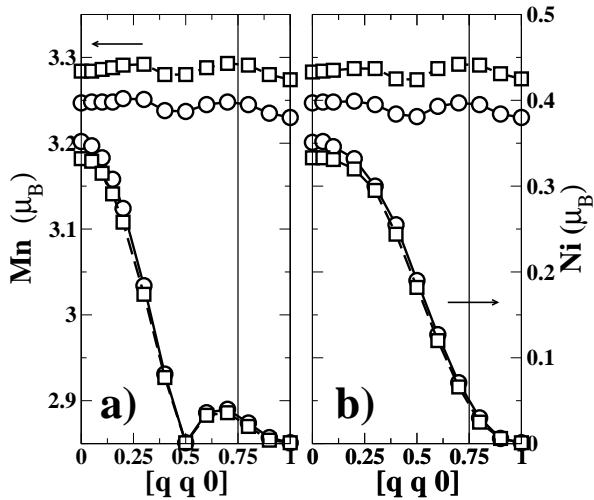}
\caption{Magnetic moments within the atomic spheres as a function of the
spiral vector $\bm{q}$. $\bigcirc$ Ni$_2$MnAl, $\square$ Ni$_2$MnGa. a) Ni
atoms are equivalent, b) Ni atoms are inequivalent}
\label{fig:mom1}
\end{figure}

\begin{figure}[h]
\centering \epsfig{file=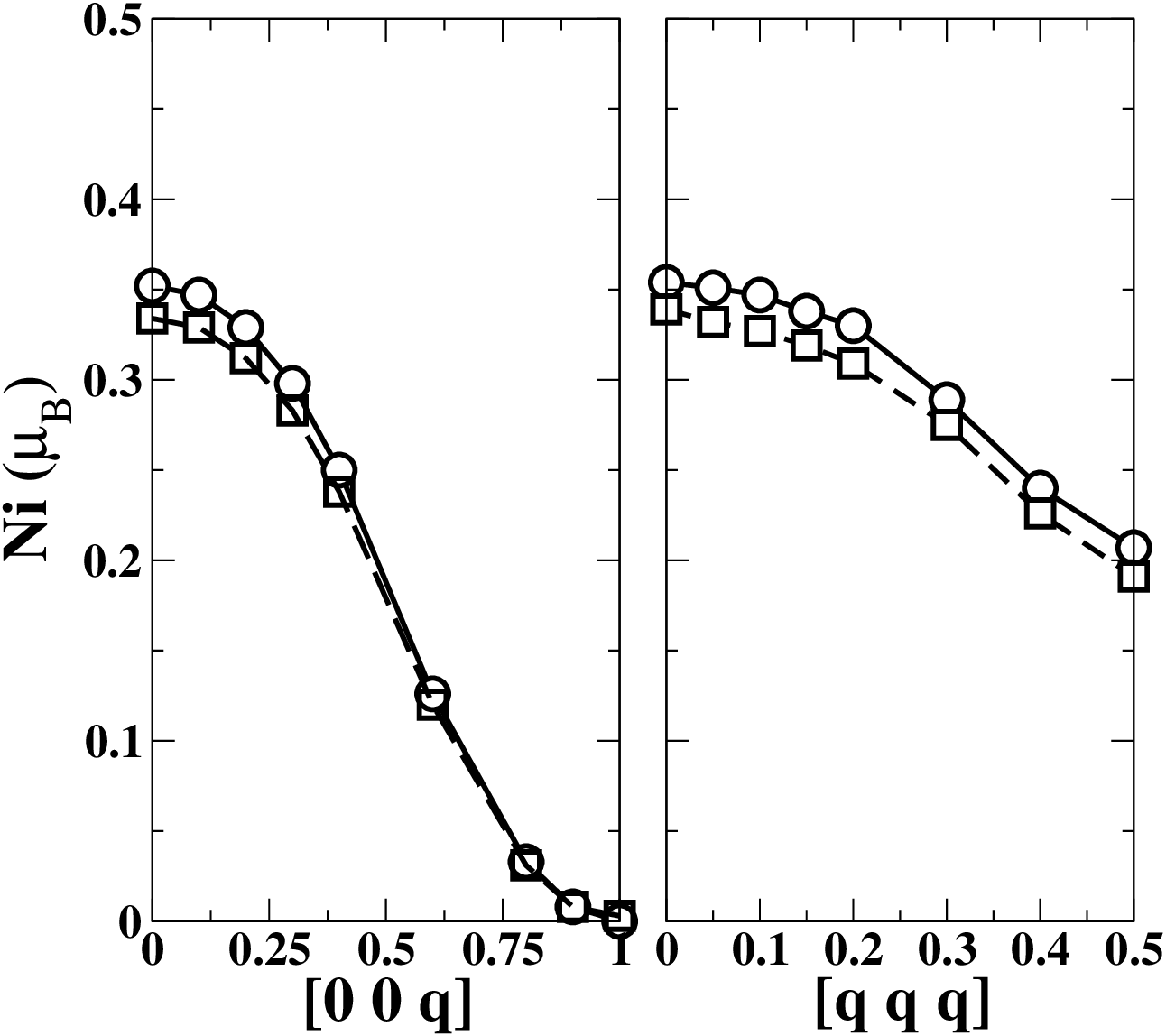,width=.9\columnwidth,keepaspectratio=true}
\caption{Magnetic moments within the Ni sphere as a function of the spiral vector $\bm{q}$. $\bigcirc$ Ni$_2$MnAl, $\square$ Ni$_2$MnGa.}
\label{fig:mom2}
\end{figure}

In order to obtain a deeper understanding of the energy dispersion we next
look into the behaviour of magnetization. The
magnetic moments averaged over the atomic spheres for different $\bm{q}$ are
shown in Figs.~\ref{fig:mom1}~and~\ref{fig:mom2}.
The atomic magnetizations show that within the Mn spheres the magnetization is
nearly constant and the variation in the total magnetization is mainly due to
Ni. Also, the symmetry consideration of the equivalence of Ni atoms has no effect
on the Mn moment. This points to a more localized character of the magnetic
moment 
of Mn, 
compared to a more itinerant character of Ni. Because most 
of the total magnetic moment comes from Mn, these alloys can be considered as
localized-moment systems consistent with the
traditional view for similar materials 
\cite{KublerEtal83}. However, despite the relative smallness of its magnetic
moment, Ni has significant effect for the energetics as discussed later on. The
differences between Ni$_2$MnGa and 
Ni$_2$MnAl are small: the magnetic moment in Ni$_2$MnGa is slightly larger
as shown already in previous work \cite{Ayuela99}.

As the magnetic moment in Ni shows a larger variation, the behaviour of the Ni 
moment 
is analyzed in more detail for several directions. 
The magnetization decreases monotonously both in the [001] and in the [111]
directions. Differences are at the antiferromagnetic configurations as the
magnetic moment of Ni remains finite at $\bm{q}=(1\ 1\ 1)$ but vanishes at
$\bm{q}=(0\ 0\ 1)$. 
In the [110] direction, the behaviour of the Ni moment depends strongly on the
symmetry as seen in Fig.~\ref{fig:mom1}. When the two magnetic moments are
forced to be the same, the  magnetization starts to decrease with
increasing $q$ and 
vanishes to zero value at $\bm{q}=(0.5\ 0.5\ 0)$. For larger $q$ values the
moment shows a small peak before 
decreasing again to zero in the antiferromagnetic state at $\bm{q}=(1\ 1\
0)$. In the case of Ni atoms being inequivalent only a monotonous decrease
similar to the [001] direction is seen. 

Because most of the variation in total magnetization is due to Ni, it should
have a larger effect also on the energy dispersion. The importance of Ni can
be seen most clearly in the [110] direction for the cases of different
symmetry. The symmetry affects only Ni as seen in the behaviour of the
magnetization, Fig.~\ref{fig:mom1}. As the energy dispersion depends on the
symmetry, Fig.~\ref{fig:ene2}, the importance of Ni is clear. It can be noted
also that the energy lowers when the Ni moment increases.  
Based on the above reasoning, Ni
should have an effect on the Curie temperature, which indeed is seen in
experiments 
where the increase in Ni content decreases the Curie temperature
\cite{Vasilev99}.   

\begin{figure}[h]
\centering
\epsfig{file=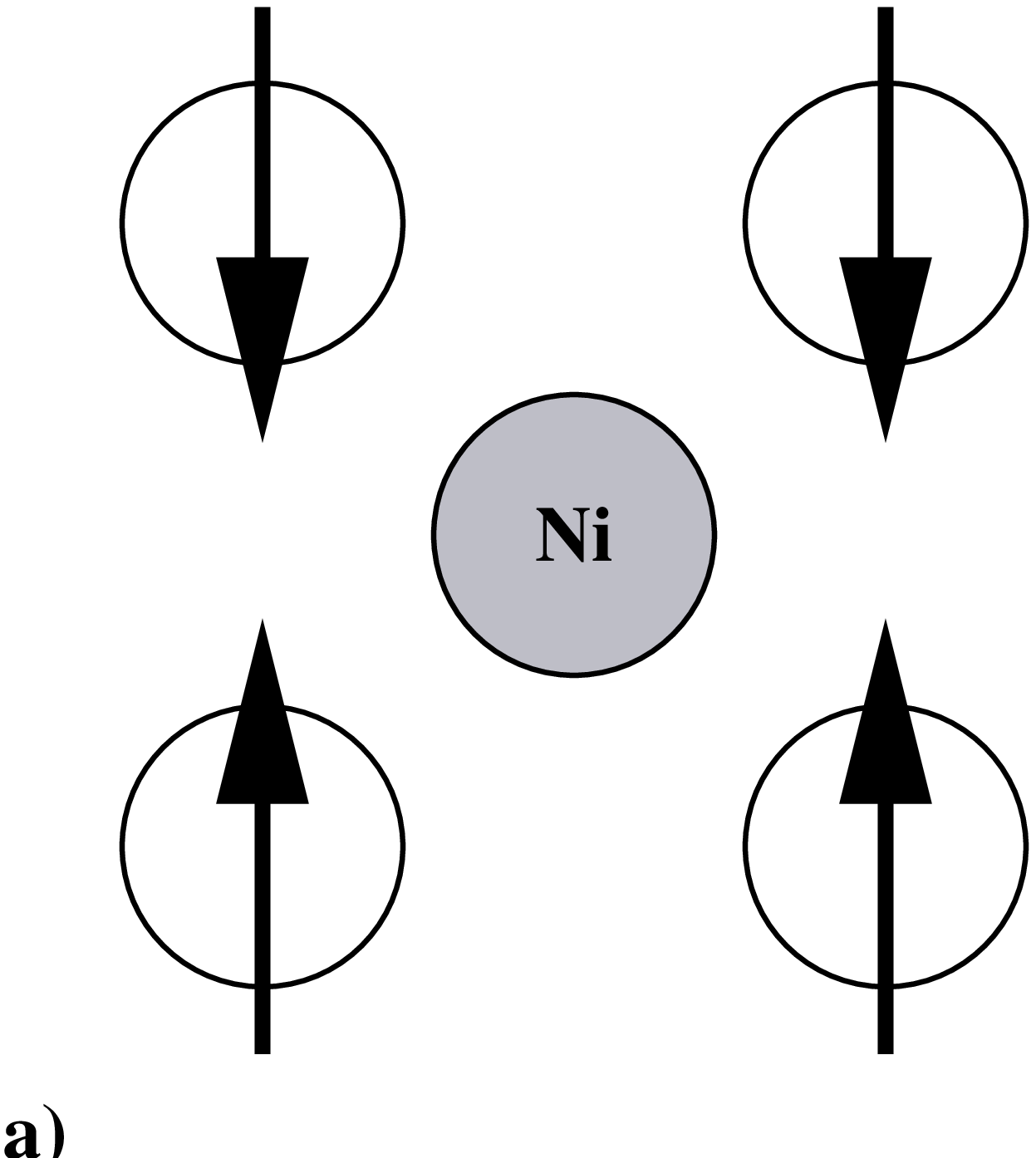,width=.36\columnwidth,keepaspectratio=true}
\hspace{.12\columnwidth}
\epsfig{file=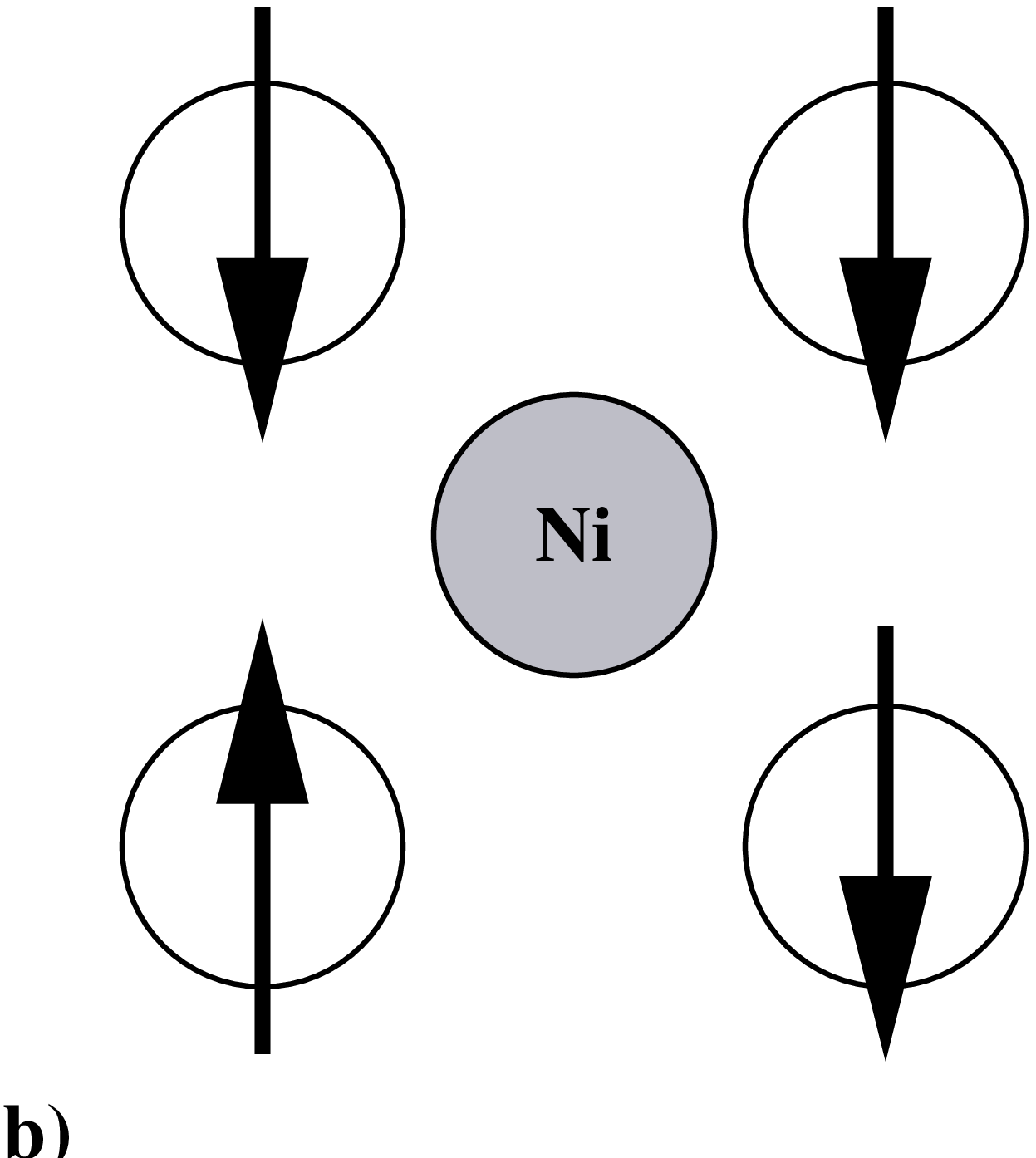,width=.36\columnwidth,keepaspectratio=true} 
\caption{Schematic view of the magnetic moments in the nearest neighbour Mn
atoms of Ni at a) $\bm{q}=(0\ 0\ 1)$ b) $\bm{q}=(0.5\ 0.5\ 0.5)$.}  
\label{fig:sche1}
\end{figure}

The variation of the Ni moment can be understood by
considering symmetry arguments and the coordination around Ni atoms. In
the [001] direction two 
of the four Mn atoms neighbouring Ni have the same magnetization direction in
the spiral and the other two have different direction, as shown schematically
in Fig.~\ref{fig:sche1}(a). The magnetization in Ni
favours ferromagnetic alignment with the neighbouring Mn moments so that part
of the Ni moment can be thought to align with one group of the Mn neighbours
and part with the other group. The total moment within the atomic sphere is
average of these two parts and the Ni moment decreases when the
angle between the Mn moments increases. In the antiferromagnetic configuration
there is a complete frustration of the Ni atoms which results in the zero average
magnetization within the sphere. For the [111] direction shown in
Fig.~\ref{fig:sche1}(b)   
one group contains three 
Mn atoms and the other group only one. Therefore the variation of the average
moment in the Ni 
sphere is smaller and the moment remains finite in the
antiferromagnetic configuration. 

\begin{figure}[h]
\centering
\epsfig{file=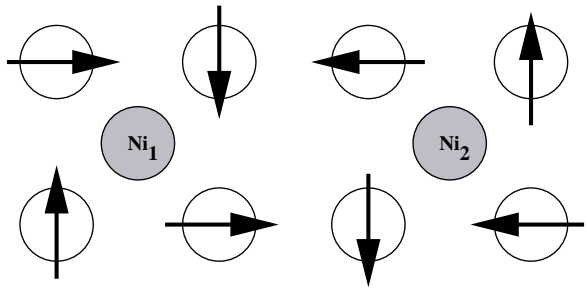,width=.9\columnwidth,keepaspectratio=true}
\caption{Schematic view of the magnetic moments in the nearest neighbour Mn
atoms of the two equivalent Ni at $\bm{q}=(0.5\ 0.5\ 0)$. }
\label{fig:sche2}
\end{figure}

In the [110] direction the situation is more
complex especially when the two Ni atoms are treated as equivalent. In the
antiferromagnetic configuration the coordination is  
similar to the case of the [001] direction. There are two groups of neighbouring
Mn with antiparallel magnetization, and the frustration leads to zero average
moment within the Ni sphere. The Ni moment
is, however, zero also at $\bm{q}=(0.5\ 0.5\ 0)$. At this point there are three
groups of equivalent Mn neighbours. One group contains two Mn atoms and the other groups 
contain one Mn atom. The magnetic moments of single Mn atoms are antiparallel to each other
and have $90^\circ$ angle with
respect to the moments in the group of the two Mn atoms. 
The other equivalent Ni atom has
three similar groups of neighbouring Mn as the first Ni. The important point is 
that the moments in the group with two Mn atoms are antiparallel to those in
the corresponding group of the first Ni, as seen in
Fig.~\ref{fig:sche2}. There is now frustration for Ni, but  
only when both equivalent Ni atoms and their neighbours are taken into
account. This frustration causes the magnetic moment around Ni to vanish
completely in contrast to the antiferromagnetic case, where a small moment
remains near Ni but averages to zero. When the Ni are inequivalent, they can
relax according the local environment so that a finite moment can remain at
$\bm{q}=(0.5\ 0.5\ 0)$ 

\begin{figure}[h]
\centering \epsfig{file=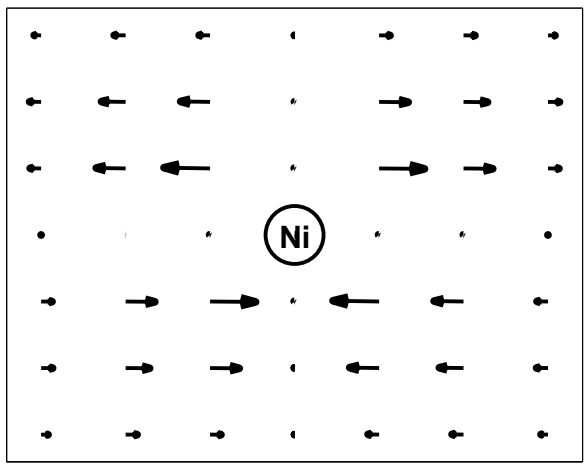,width=.9\columnwidth,keepaspectratio=true}
\caption{Magnetization density around Ni in the (001) plane with
$\bm{q}=(1\ 1\ 0)$. The width and the height of the area are 2.5 a.u. while
magnetization is in arbitrary units.}
\label{fig:magndens}
\end{figure}

An example of the magnetization density for the case where finite magnetic
moments near the Ni atom average to zero is seen in
Fig.~\ref{fig:magndens}. Here the  magnetization direction can change its sign 
within the atomic sphere. This finding shows the importance of the full
magnetization treatment when dealing with several magnetic sublattices.  

\section{Conclusions}
\label{sec:conclusions}

We have studied non-collinear magnetic configurations in the ternary alloys
Ni$_2$MnGa and  
Ni$_2$MnAl with first-principles calculations. The calculations show that
the magnetic properties are similar for both materials. The ferromagnetic
configuration is the ground state in the L2$_1$ structure, so that the
experimentally observed antiferromagnetism of Ni$_2$MnAl is related to
structural disorder.
The calculated total
energies are used to estimate the spin stiffness constant and the Curie
temperature which are in good 
agreement with the experiments. 
The similarity in the energy dispersion for
both materials suggests that the Curie temperatures should be also similar.
In the [110] direction Ni$_2$MnAl has higher energy, so that the Curie
temperature should be a slightly higher. 

The variation of the magnetic moment in the spirals shows that the Mn moment
is nearly constant while the Ni moment varies strongly. The symmetry of the
spin spiral constrains the direction of magnetization, and as Ni favours
ferromagnetic coupling with Mn, there can be frustration at certain wave
vectors resulting in the vanishing of the magnetic moment near the Ni
sites. It is also shown how there can be strong variation in the direction of
the magnetization near the atomic sites which points to the relevance of the
full magnetization treatment.  

Some conclusions can be made concerning the role of the constituent atoms for
the 
magnetic properties. As the magnetic moment of Ni varies strongly and its
symmetry affects the energy considerably,
Ni has probably a strong effect on the energy dispersion
especially when larger wave vectors are involved. Therefore Ni also
influences 
the Curie temperature. If one assumes that the spin stiffness is mainly due to
Mn, and the lowering of the energy with larger wave vectors due to Ni, Ni
lowers the Curie temperature from 830 K to 485 K within the present
approximations. As the increase in the Ni moment decreases the energy 
it is suggested that in order to increase the Curie temperature
one should replace some Ni, perhaps a little counter-intuitively, with 
some non-magnetic element, for example Cu. Further experiments should clarify 
these  issues and confirm the above suggestions.

\begin{acknowledgments}
This work has been supported by the Academy of Finland
(Centers of Excellence Program 2000-2005) and by the National Technology
Agency of Finland (TEKES) and the consortium of Finnish companies (ABB
Corporate Research Oy, AdaptaMat Oy, Metso Oyj, Outokumpu Research
Oy). A. Ayuela has been supported by the EU TMR program (Contract
No. ERB4001GT954586 ). Computer facilities of the Center for 
Scientific Computing (CSC) Finland are greatly acknowledged.
L. N. acknowledges the supports from the Swedish Research Council
and the Swedish Foundation for Strategic Research.
\end{acknowledgments}

\bibliographystyle{apsrev}
\bibliography{omaabbrev,erik2,dft,paperbib}

\end{document}